\documentclass[12pt,letterpaper,final]{iopart}

\usepackage{hyperref}
\usepackage{graphicx}
\usepackage[margin=1in]{geometry}

\expandafter\let\csname equation*\endcsname\relax
\expandafter\let\csname endequation*\endcsname\relax
\usepackage{amsmath}
\usepackage{amssymb}

\newcommand{\beq}{\begin{equation}}
\newcommand{\eeq}{\end{equation}}
\newcommand{\bea}{\begin{eqnarray}}
\newcommand{\eea}{\end{eqnarray}}

\begin{document}

\title{Tuning the Fermi level through the Dirac point of giant Rashba semiconductor BiTeI with pressure}

\author{D.\ VanGennep$^1$, S.\ Maiti$^{1,2}$, D.\ Graf$^2$, S.\ W.\ Tozer$^2$, C.\ Martin$^{1,3}$, H.\ Berger$^4$, D.\ L.\ Maslov$^1$, J.\ J.\ Hamlin$^1$}

\address{$^1$Department of Physics, University of Florida, Gainesville, FL 32611\\
	  $^2$National High Magnetic Field Laboratory, Tallahassee, FL 32310\\
	  $^3$School of Theoretical and Applied Science, Ramapo College of New Jersey, Mahwah, NJ 07430\\
	  $^4$Institute of Condensed Matter Physics, \'{E}cole Polytechnique F\'{e}d\'{e}rale de Lausanne, CH-1015 Lausanne, Switzerland}
\ead{jhamlin@ufl.edu}
\begin{abstract}
We report measurements of Shubnikov-de Haas oscillations in the giant Rashba semiconductor BiTeI under applied pressures up to $\sim 2\,\mathrm{GPa}$.  We observe one high frequency oscillation at all pressures and one low frequency oscillation that emerges between $\sim 0.3-0.7\,\mathrm{GPa}$ indicating the appearance of a second small Fermi surface. BiTeI has a conduction band bottom that is split into two sub-bands due to the strong Rashba coupling, resulting in a `Dirac point'. Our results suggest that the chemical potential starts below the Dirac point in the conduction band at ambient pressure and moves upward, crossing it as pressure is increased. The presence of the chemical potential above this Dirac point results in two Fermi surfaces. We present a simple model that captures this effect and can be used to understand the pressure dependence of our sample parameters. These extracted parameters are in quantitative agreement with first-principles calculations and other experiments. The parameters extracted via our model support the notion that pressure brings the system closer to the predicted topological quantum phase transition.
\end{abstract}

\maketitle

\section{Introduction}
\pagestyle{plain}
In materials that combine broken inversion symmetry with a large spin-orbit coupling, the so-called ``Rashba effect'' can cause lifting of the spin-degeneracy leading to two chiral sub-bands.  Such strong spin-orbit coupled materials have potential use in spintronic applications~\cite{marchenko_2012_1,takayama_2012_1}.  Until recently, systems exhibiting large Rashba splittings were restricted largely to surfaces~\cite{lashell_1996_1,koroteev_2004_1}, interfaces~\cite{nitta_1997_1}, and thin films~\cite{dil_2008_1,hirahara_2006_1,hirahara_2007_1}.  BiTeI appears to exhibit a very large Rashba splitting of the \textit{bulk} electronic bands~\cite{ishizaka_2011_1} and, for this reason, has attracted significant attention.

First-principles calculations have predicted that, under presssure, BiTeI undergoes a band inversion and becomes the first instance of a non-centrosymmetric topological insulator~\cite{bahramy_2011_1}.  The critical pressure, $P_c$, for the band inversion is predicted to be in the range of 1.7 - 4.1 GPa.  Infrared reflectance and transmission measurements under pressure found signatures of the predicted topological quantum phase transition (TQPT)~\cite{xi_2013_1}.  In particular, consistent with the approach to a linear dispersion near $P_c$, a maximum in the plasma frequency, $\omega_p$ (and hence the the Fermi velocity, $\nu _F$) was found near 2.2 GPa.  X-ray diffraction measurements, also reported in that study, found that the crystal structure remained unchanged from the ambient pressure one below $\sim 9\,\mathrm{GPa}$, although the ratio of the lattice constants, $c/a$, passes through a weak minimum near $P_c$~\cite{xi_2013_1,chen_2013_1}.  Another infrared study under pressure~\cite{tran_2014_1} did not report clear evidence for the TQPT, though in that work the number of pressures sampled near $P_c$ was lower.  A combined theoretical and high pressure x-ray diffraction study suggested that the band inversion occurs closer to 4.5~GPa~\cite{chen_2013_1}.  In order to further address the effects of compression on BiTeI, we have carried out measurements of Shubnikov-de Haas (SdH) oscillations under pressure.

A number of recent papers have reported measurements of SdH oscillations in BiTeI at ambient pressure~\cite{martin_2013_1,bell_2013_1,murakawa_2013_1}.  Although BiTeI is nominally a semiconductor with a band gap of $\sim 0.3\,\mathrm{eV}$, self doping (\textit{i.e.} non-stoichiometry) leads to a chemical potential, $\mu$, that lies in the conduction band, producing carrier densities on the order of $\sim 10^{19}\,\mathrm{cm^{-3}}$. The samples examined in these studies fall into two categories, with the $\mu$, lying either above or below the Dirac point that arises at the crossing point of the two chiral sub-bands.  For the case where $\mu$ lies below the Dirac point, only the so-called outer Fermi surface (OFS) is present, and only a single oscillation frequency, $F^-$, is observed.  For cases where $\mu$ lies above the Dirac point, an inner Fermi surface (IFS) is present and, consequently, a second oscillation frequency, $F^+$, is also observed.  In this paper, we report measurements of Shubnikov-de Haas oscillations in BiTeI under applied pressures up to 1.9 GPa.  These measurements show that, for the sample examined in this study, $\mu$ appears to lie at or below the Dirac point at ambient pressure, and is pushed above it with increasing pressure.

\section{Experimental methods}
Single crystals of BiTeI were grown by the chemical vapor transport method.  A small piece of sample with dimensions of about $700\,\mathrm{\mu m} \times 200\,\mathrm{\mu m} \times 40\,\mathrm{\mu m}$ was cut from a larger crystal.  Pt wires were attached to the crystal using Dupont 4929N conductive silver paste.  The sample was mounted to the wire and fiber optic feed-through of a piston-cylinder pressure cell which was constructed from MP35N alloy. The pressure was calibrated at room temperature and again at the lowest temperature using the fluorescence of the R1 peak of a small ruby chip~\cite{piermarini_1975_1}. Daphne 7474 oil was used as the pressure-transmitting medium~\cite{murata_2008_1} surrounding the sample. The resistance was measured using a Quantum Design PPMS for each pressure up to 1.9 GPa. A Lakeshore 370 resistance bridge was used for the transport measurement.  Four wire electrical resistivity measurements were carried out between 2-300 K, in magnetic fields up 16 tesla, at pressures of 0.3, 0.7, 1.1, 1.4, and 1.9 GPa.  The sample resistance was measured in the crystalline $ab$-plane, while the magnetic field was applied parallel to the $c$-axis.

\section{Results}
\begin{figure}[p]
  \raggedleft
    \includegraphics[width=0.85\textwidth]{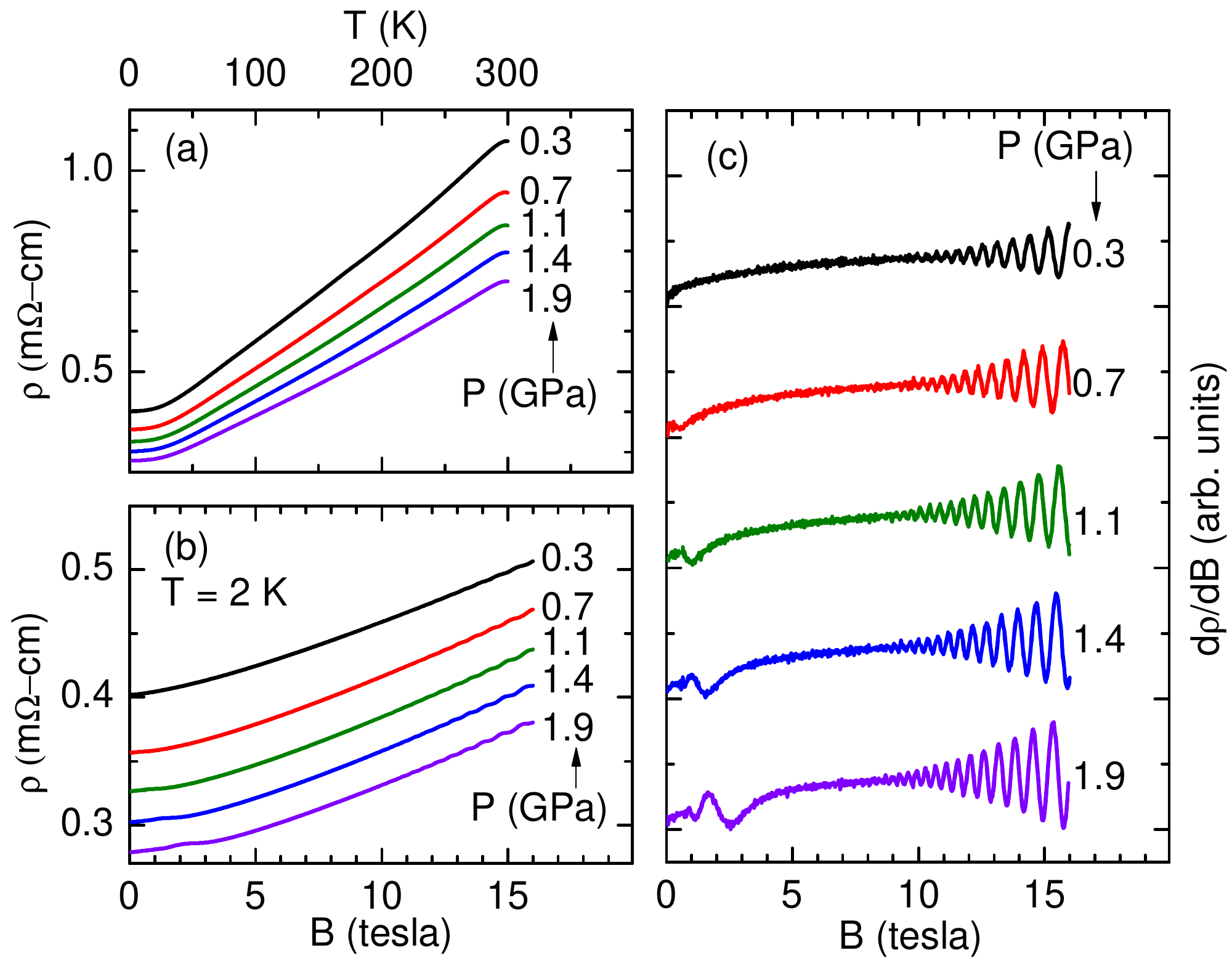}
    \caption{(a) Electrical resistivity in the $ab$-plane versus temperature, measured from 2-300 K at several pressures. (b) Resistivity versus magnetic field for several pressures.  The magnetic field was applied parallel to the $c$-axis.  At the lowest pressures, only one set of oscillations is visible (above $\sim 12\,\mathrm{tesla}$).  At higher pressures, a second set of oscillations emerges below $\sim 3\,\mathrm{tesla}$. (c) These features are more clearly visible in a plot of the derivative of the resistivity with respect to field.  The derivative plots have been offset vertically for clarity.}
 \label{fig1}
\end{figure}

Fig.~\ref{fig1} presents on overview of our results.  Increasing pressure leads to a slight suppression in the overall magnitude of the resistivity, although the shape of the curves do not change significantly (Fig.~\ref{fig1}a).  Fig.~\ref{fig1}b shows the magnetic field dependence of the resistivity for each pressure.  For all pressures, oscillations of the resistivity are visible in the raw data.  The oscillations can be more clearly observed by plotting the field derivative of the resistivity, as shown in Fig.~\ref{fig1}c.  Plotting the data in this way reveals the presence of two regimes.  At low pressures, only one set of oscillations is apparent, evident at fields above $\sim 10\,\mathrm{tesla}$.  Beginning at pressures above $\sim 1\,\mathrm{GPa}$, a second, lower frequency oscillation appears at low fields.  As pressure increases the low field oscillations become larger in amplitude and persist to higher fields.  At the highest pressure the low field oscillations die-out for fields above about $3\,\mathrm{tesla}$.

The oscillations at $1.9\,\mathrm{GPa}$ closely resemble those observed at ambient pressure in Ref.~\cite{murakawa_2013_1}.  The higher frequency oscillations have been identified as being associated with the OFS while the low frequency oscillations are associated with the IFS.  The absence of the IFS oscillation at the lowest pressure and it's appearance under pressure, suggests that the chemical potential lies at or slightly below the Dirac point and ambient pressure and then rises above the Dirac point as pressure increases.  This is discussed in further detail in Section~\ref{sec:Discussion}.

\begin{figure}[p]
  \raggedleft
    \includegraphics[width=0.85\textwidth]{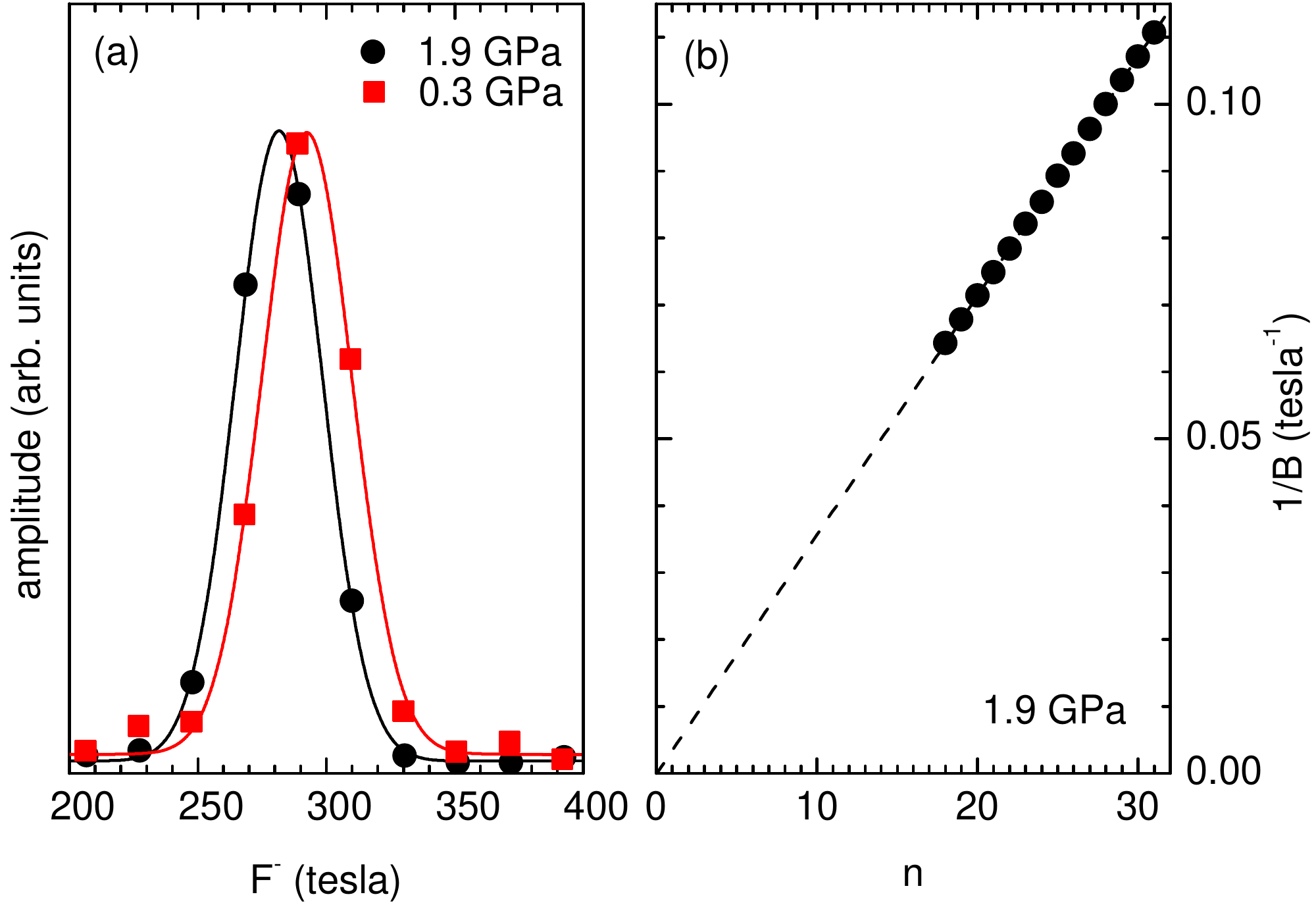}
    \caption{Data for oscillations arising from the OFS.  (a) FFT amplitude versus frequency at 0.3 and 1.9 GPa.  A small decrease in the frequency is evident. (b) Inverse field, $1/B$, versus Landau index, $n$, for the OFS at 1.9 GPa.}
 \label{fig2}
\end{figure}

\begin{figure}[p]
  \raggedleft
    \includegraphics[width=0.85\textwidth]{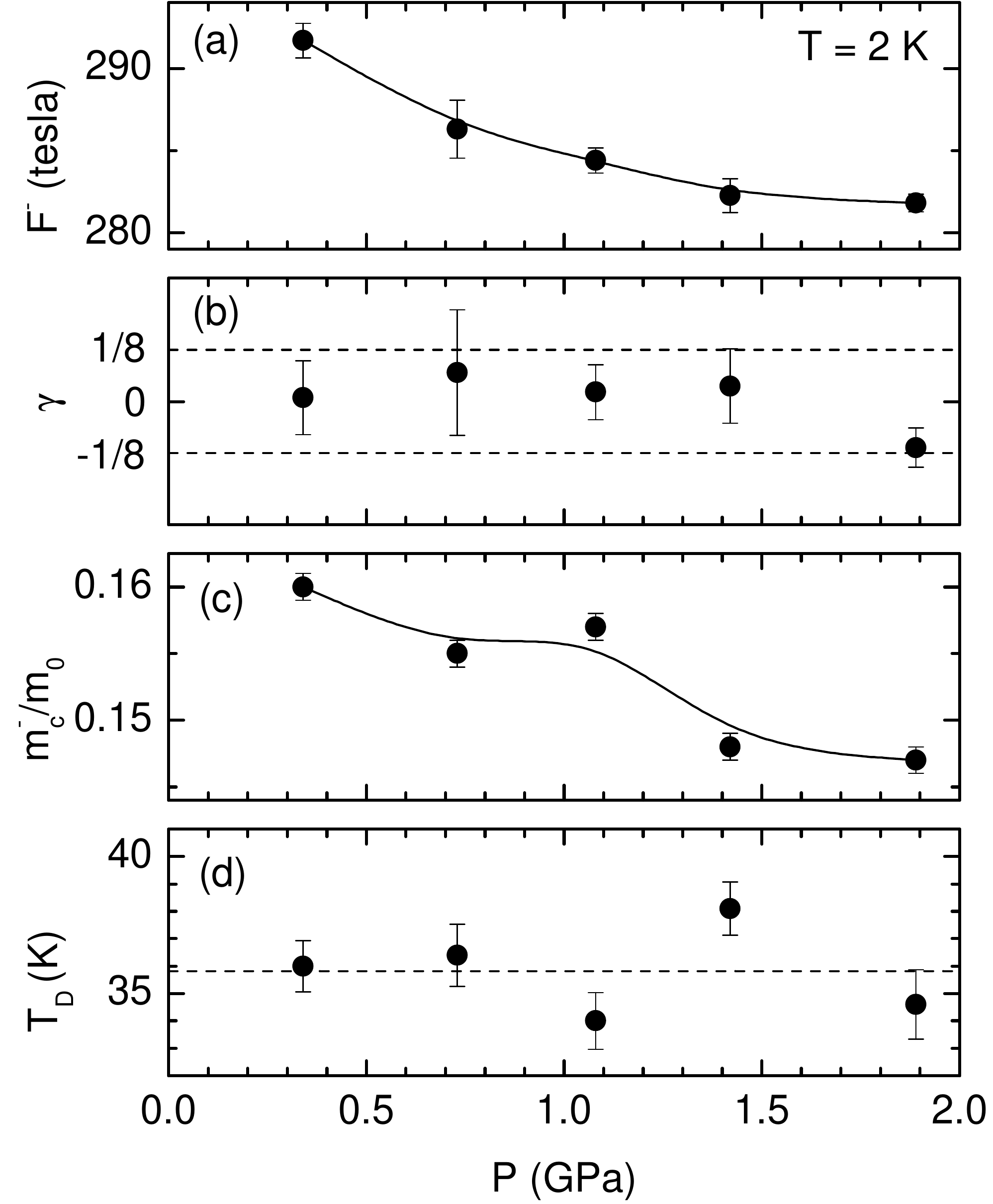}
    \caption{Parameters of the OFS at 2~K as a function of pressure.  (a) The frequency of the oscillations drops slightly, (b) the extrapolated $\gamma$ remains between $\pm 1/8$, (c) the cyclotron mass, $m_C^-$, drops slightly with pressure, (d) and the Dingle temperature, $T_D$, is roughly pressure independent.  The curved lines in (a) and (c) are guides to the eye.}
 \label{fig3}
\end{figure}

We first discuss the OFS oscillations, which are presented in Figs.~\ref{fig2} and \ref{fig3}.  The oscillations were analyzed by subtracting a polynomial background from the magneto-resistance.  The frequency was then extracted from the resulting data using a fast Fourier transform (FFT).  Fig.~\ref{fig2}a plots the normalized FFT amplitude versus frequency for the highest and lowest pressures in our experiments.  The frequency has also been determined by taking the slope of a plot of the Landau level index, $n$, versus the value of $1/B$ where the oscillatory part of the magneto-resistance passes through a maximum (see Fig.~\ref{fig2}b).  Both methods of extracting the frequency yield nearly identical results.  The frequencies extracted via index plot are plotted versus pressure in Fig.~\ref{fig3}a. A  small reduction in the frequency with pressure is evident.  Because the maximum fields in these experiments are far from the quantum limit for the OFS, we have shifted the values of $n$ by the integer value that produces the smallest intercept. The values of the intercept, $\gamma$ are plotted versus pressure in Fig.~\ref{fig3}b.  The intercept values all lie between the values of $\pm 1/8$, which is consistent with a non-trivial Berry's phase as discussed in Ref.~\cite{murakawa_2013_1}.  The temperature and field dependences of the oscillation amplitudes were used to extract the effective cyclotron mass, $m_c^-$, and Dingle temperature, $T_D$, respectively, as described in Ref.~\cite{martin_2013_1}.  Figs.~\ref{fig3}a and \ref{fig3}b show the effective mass and Dingle temperature, respectively, as a function of pressure.  The effective mass shows a weak suppression with pressure.  The lack of a strong reduction in the Dingle temperature supports the expectation that good hydrostatic conditions are maintained in the cell over the entire pressure range studied.

\begin{figure}[p]
  \raggedleft
    \includegraphics[width=0.85\textwidth]{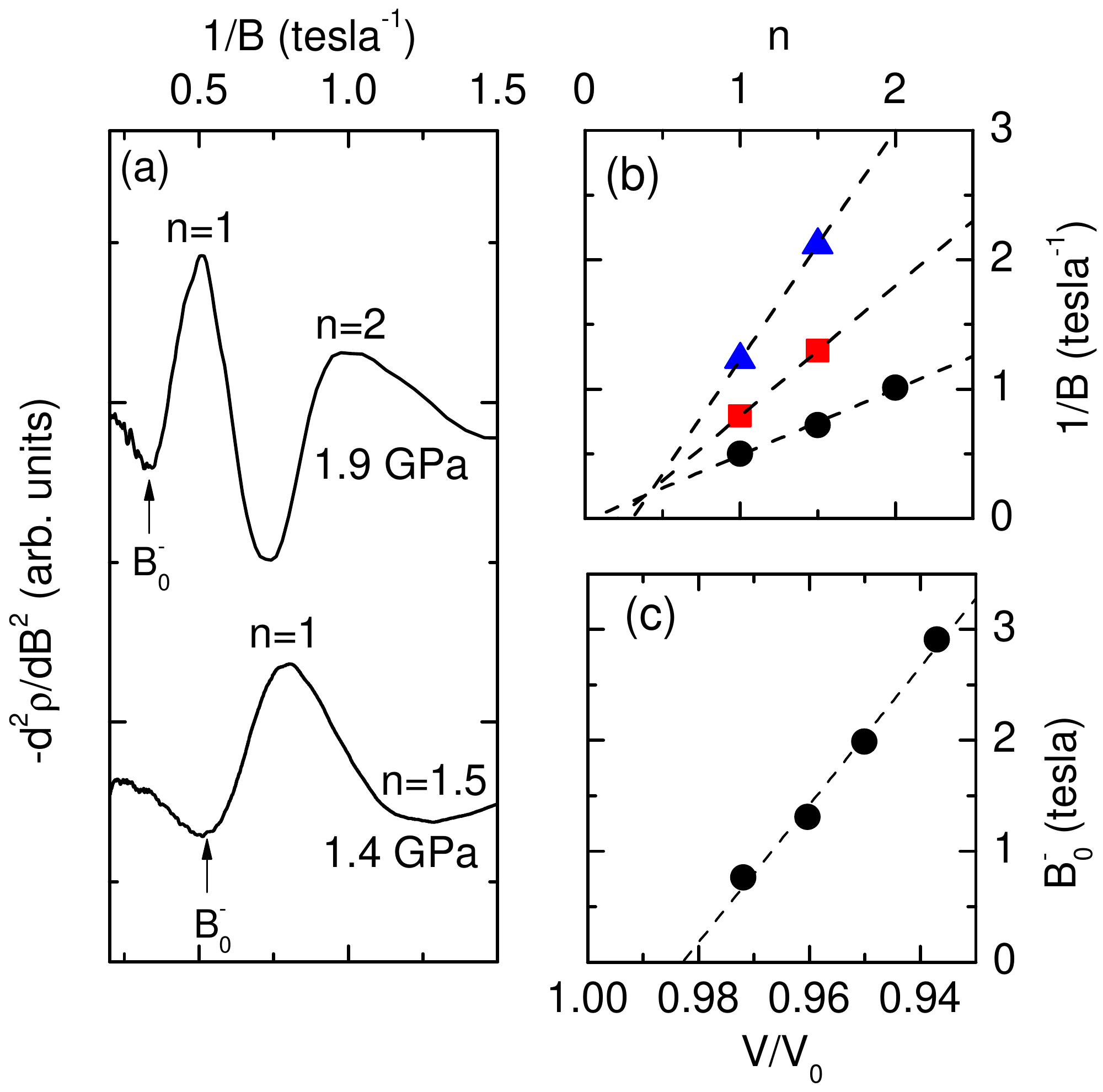}
    \caption{Data concerning oscillations arising from the IFS. (a) Negative second derivative of the resistivity with respect to field. The data have been offset vertically for clarity.  The small amplitude of the the oscillations only allows the first $1-2$ oscillations to be resolved. $B^-_0$ indicates the last dip in the data as the IFS reaches the quantum limit.  (b) Index plot for the IFS at 1.9 GPa (black circles), 1.4 GPa (red squares), and 1.1 GPa (blue triangles).  (c) A plot of $B^-_0$ versus relative volume extrapolates to zero at $V/V_0$ = 0.983.  This implies that the IFS does not appear until $\sim 0.4\,\mathrm{GPa}$}
 \label{fig4}
\end{figure}

Next, we turn to the low frequency oscillations associated with the IFS.  Fig.~\ref{fig4}a shows the negative second derivative of the resistivity, $-d^2\rho/dB^2$ versus the inverse field, $1/B$, in the vicinity of the IFS oscillations for the highest two pressures.  The Landau indices are indicated by the integer numbers located at the peaks in $-d^2\rho/dB^2$ and half integer values located at the valleys~\cite{rhoxx_note}.  For these oscillations, the quantum limit is reached below 3~tesla, presumably because of a very low carrier density.  The position of the last dip in the data before the IFS oscillations die out is denoted $B_0^-$. Because the IFS oscillations die out very quickly at low fields, it is only possible to reliably extract the $1/B$ values for the first $1-2$ Landau levels.  Nonetheless, this is enough to provide an estimate of the frequency of these oscillations.  Fig.~\ref{fig4}b plots the Landau index versus $1/B$ for the pressures where these oscillations are present.  The frequency of these oscillations is determined from the slope of a linear fit to the data and listed for each pressure in Table~\ref{tab1}.  The substantial increase in this frequency is consistent with pressure pushing the Fermi level upwards relative to the Dirac point (see Section~\ref{sec:Discussion}).  In order to obtain an estimate of the pressure where the IFS first appears, we have also plotted the $B_0^-$ feature as a function of relative volume, $V/V_0$ in Fig.~\ref{fig4}c. The applied pressures can be converted into relative volume using the equation of state determined via high-pressure x-ray diffraction measurements~\cite{xi_2013_1}.  A linear extrapolation of $B_0^-$ suggests that the IFS appears for relative volumes below 0.983, which amounts pressures above 0.4 GPa.  Using the background subtraction method described in Ref.~\cite{murakawa_2013_1} and the temperature dependent amplitude of these oscillations, we were able to arrive at an estimate of the cyclotron mass, $m_C^+/m_0 \sim 0.012$, at 1.9 GPa, for the carriers on the IFS.  The smaller amplitude of the oscillations at lower pressures prohibits a reliable estimate of the mass at lower pressures.

\section{Discussion}
\label{sec:Discussion}
We have already mentioned that the appearance of a second oscillation is consistent with an increase of the chemical potential with pressure.  In order to obtain quantitative estimates of the position of the chemical potential relative to the Dirac point, we have taken two complementary approaches.  In the first approach, we extracted band structure parameters including the band mass, the Rashba parameter $\alpha$, and chemical potential from our data using a simple model Hamiltonian.  In the second approach, we determined the chemical potential by relating our data to previous first principles calculations.  We find that both approaches yield nearly identical pressure dependences of the chemical potential.

\subsection{Analysis in terms of the Rashba Hamiltonian}
Noting that the chemical potential lies somewhere in the vicinity of $\sim 100$ meV from the bottom of the conduction band~\cite{bahramy_2011_2,ishizaka_2011_1,murakawa_2013_1,martin_2013_1,bell_2013_1} we model the bottom of the conduction band by the usual Rashba Hamiltonian:
\beq
  \mathcal{H}_k = \hbar^2\left(\frac{k_x^2+k_y^2}{2m_1}+\frac{k_z^2}{2m_2}\right)\hat{I}
  + \alpha(\mathbf{\sigma}\times \mathbf{k})_z - \mu
\eeq
where $m_{1,2}$ are the effective in-plane ($m_1$) and out of plane ($m_2$) band masses, $\alpha$ is the Rashba parameter that encodes the strength of the spin-orbit interaction, $\mu$ is the chemical potential, $\mathbf{\sigma}$ is the vector of Pauli matrices, and the $z$-axis is along the normal to the $a-b$ planes. We have added the $k_z$ dispersion to reproduce the Fermi surface topology in 3D~\cite{bell_2013_1,murakawa_2013_1}.  The $k_z$ dispersion is not significant for our study as we apply the magnetic field along along the $z$-direction. Thus, $k_z$ remains a good quantum number and only the value of $k_z$ corresponding to the maximum cross-sectional area of the Fermi surface ($k_z=0$ in this model) is relevant. The magnetic field quantizes the $x-y$ motion into Landau levels given by~\cite{bychkov_1984_1}:
\beq
\label{eq:LandauLevels}
E^{\pm}_n = \hbar \omega_c \left(n\pm\sqrt{\frac{4E_R}{\hbar\omega_c}n + \Delta^2}\right),~~~n\ge1,\\
\eeq
where `+' and `-' correspond to the inner and outer Rashba sub-bands and the Zeeman term, $\Delta$, is given by $\Delta = \frac12\left(1-\frac{gm_1}{2m_0}\right)$, where $g$ is the effective $g$-factor, $\omega_c = \frac{Be}{m_1}$, and $m_0$ is the bare mass of the electron. Within this model, it is natural to define a Rashba energy scale $E_R = \frac12m_1\left(\frac{\alpha}{\hbar}\right)^2$, which measures the position of the Dirac point relative to the bottom of the conduction band.  The three relevant band structure parameters of this model are $\mu$ (measured from the Dirac point), $\alpha$, and $m_1$ ($m_2$ is not relevant for this study). All of these parameters can be extracted from our measurements at each pressure and thereby one can examine how pressure affects the band structure. To relate this model to BiTeI, one simply needs to translate the origin from $k_z = 0$ to the $A$ point of the Brillouin zone~\cite{bahramy_2011_1}.

\subsubsection{Linearity of index plots:}
In the work of Murakawa~\textit{et al.}~\cite{murakawa_2013_1}, it was found that the index plots for both sub-bands were linear to the lowest Landau levels. We show below that our simplified model captures this behavior if $\Delta$ is small. To see this, we solve Eq.~(\ref{eq:LandauLevels}) for the fields at which $E_n=\mu$:
\beq
  \label{eq:Index plots2}
  \frac{m_1\mu}{\hbar e}\frac{1}{B_n^{\pm}} = n(1+2f)\pm\sqrt{4f(1+f)n^2+\Delta^2}.
\eeq
where $ f = \frac{E_R}{\mu}$ and $\mu$ is taken to be above the Dirac point.  We see that the dependence of $1/B_n$ on $n$ is, in general, non-linear.  However, if $4n^2f(1+f)\gg \Delta^2$, the dependence reduces to:
\beq
  \label{eq:Index plots3}
  \frac{m_1\mu}{\hbar e}\frac{1}{B_n^{\pm}} \approx n(\sqrt{1+f}\pm\sqrt{f})^2+\mathcal{O}\left(\frac{\Delta^2}{4n\sqrt{f(1+f)}}\right).
\eeq
Thus, the observed linearity of the index plots suggests a negligible role of the Zeeman term.  Using this input from the experiment, we drop the Zeeman terms ($\Delta$) in our subsequent analysis.

\subsubsection{Evolution of band structure parameters with pressure:}
\label{sec:evolution}
We focus on the data sets that show two frequencies. This translates to the regime with $\mu>0$, where there are two Fermi surfaces, and the two frequencies in SdH oscillations correspond to the maximal cross sectional areas of each Fermi surface. From Eq.~(\ref{eq:Index plots3}), these frequencies are found as:
\beq
  \label{eq:freq}
  F^{\pm} = \frac{\mu m_1}{\hbar e}\left(\sqrt{1+f} \mp \sqrt{f}\right)^2.
\eeq
The cyclotron masses ($m_C^{\pm}$), extracted from the temperature dependence of the SdH amplitude, are proportional to the derivatives of the areas with respect to the energy evaluated at the Fermi level:
\beq
  \label{eq:mass}
  m^{\pm}_C = m_1\left(1 \mp \frac{\sqrt{f}}{\sqrt{1+f}}\right).
\eeq
Comparing Eqs.~(\ref{eq:freq}) and (\ref{eq:mass}), we find that
\beq
  \label{eq:ratio}
  \frac{m^+_C}{m^-_C} = \sqrt{\frac{F^+}{F^-}},
\eeq
and
\beq
  \label{eq:ratios}
  \frac{F^-}{F^+} = \left(\frac{\sqrt{1+f}+\sqrt{f}}{\sqrt{1+f}-\sqrt{f}}\right)^2.
\eeq
Using these equations, it is possible to extract the parameters $m_1$, $\alpha$, and $\mu$ from the measured values of $F^{\pm}$ and $m_C^{\pm}$. Eq.~(\ref{eq:ratio}) provides a useful consistency check. 

In Table~\ref{tab1}, we report the results of this analysis on our data for the pressures where both oscillations can be observed (1.1-1.9 GPa) and, for comparison, on the ambient pressure data of Murakawa~\textit{et al.}~\cite{murakawa_2013_1}.  For our data, we used Eq.~(\ref{eq:ratio}) to estimate $m_C^+$ at 1.1 and 1.4 GPa, since, as mentioned earlier, it was not possible to reliably extract the mass at these pressures due to the small amplitude of oscillations.  At 1.9 GPa, this approach yields $m_C^+$ within $\sim 3\%$ of the measured value.  The values of the Rashba parameter, $\alpha$, presented in Table~\ref{tab1} also agree very well with those determined through ARPES measurements~\cite{ishizaka_2011_1}.  Both $\alpha$ and the band mass, $m_1$, do not appear to be strongly pressure dependent, at least over this range of pressures.  However, the chemical potential shows a clear increase with pressure.
\begin{table}
  \centering
  \begin{center}
  \begin{tabular}{l|l|l|l|l}
  & Ref.~\cite{murakawa_2013_1} & This work & This work & This work\\
  & Sample A & \mbox{1.1 GPa} & 1.4 GPa & 1.9 GPa\\ \hline
      $F^+\,\mathrm{(tesla)}$ & 3.4 & 0.6(1) & 1.0(1) & 2.0(2)\\
      $F^-\,\mathrm{(tesla)}$ & 347 & 284.4(3) & 282.3(4) & 281.8(3)\\
      $m_C^+/m_0$ & 0.023 & 0.007(1) & 0.009(1) & 0.012(1)\\
      $m_C^-/m_0$ & 0.183 & 0.157(1) & 0.148(1) & 0.147(1)\\ \hline
      $m_1/m_0$ & 0.11 & 0.082(1) & 0.078(1) & 0.080(1)\\
      $\mu\,\mathrm{(meV)}$ & 66 & 18(1) & 25(1) & 34(2)\\
      $\alpha\,\mathrm{(eV}$\AA{}) & 4.3 & 4.1(1) & 4.2(1) & 4.1(1)
 \end{tabular}
 \end{center}
 \caption{Microscopic parameters $m_1$, $\alpha$, and $\mu$ determined from frequencies, $F^{\pm}$, and cyclotron masses, $m_C^{\pm}$ taken from SdH measurements, using the method described in the text.}
 \label{tab1}
\end{table}

The black circles in Fig.~\ref{fig5}a present the dependence of $\mu$ on the relative volume $V/V_0$. A linear extrapolation of the data suggests that the chemical potential is located about $10\,\mathrm{meV}$ below the Dirac point at ambient pressure, crosses the Dirac point near $V/V_0 = 0.984$ (0.4~GPa), and rises to $\sim 34\,\mathrm{meV}$ above the Dirac point at our highest pressure (1.9~GPa).  This analysis results in the same estimate of the pressure at which the chemical potential crosses the Dirac point (0.4 GPa) as was obtained by plotting the quantum limit field, $B_0^-$, versus volume (see Figure~\ref{fig4}c). Future Hall effect measurements under pressure might be sensitive to the appearance of the IFS under pressure.

\subsection{Comparison with first principles calculations}
\begin{figure}[p]
  \raggedleft
    \includegraphics[width=0.85\textwidth]{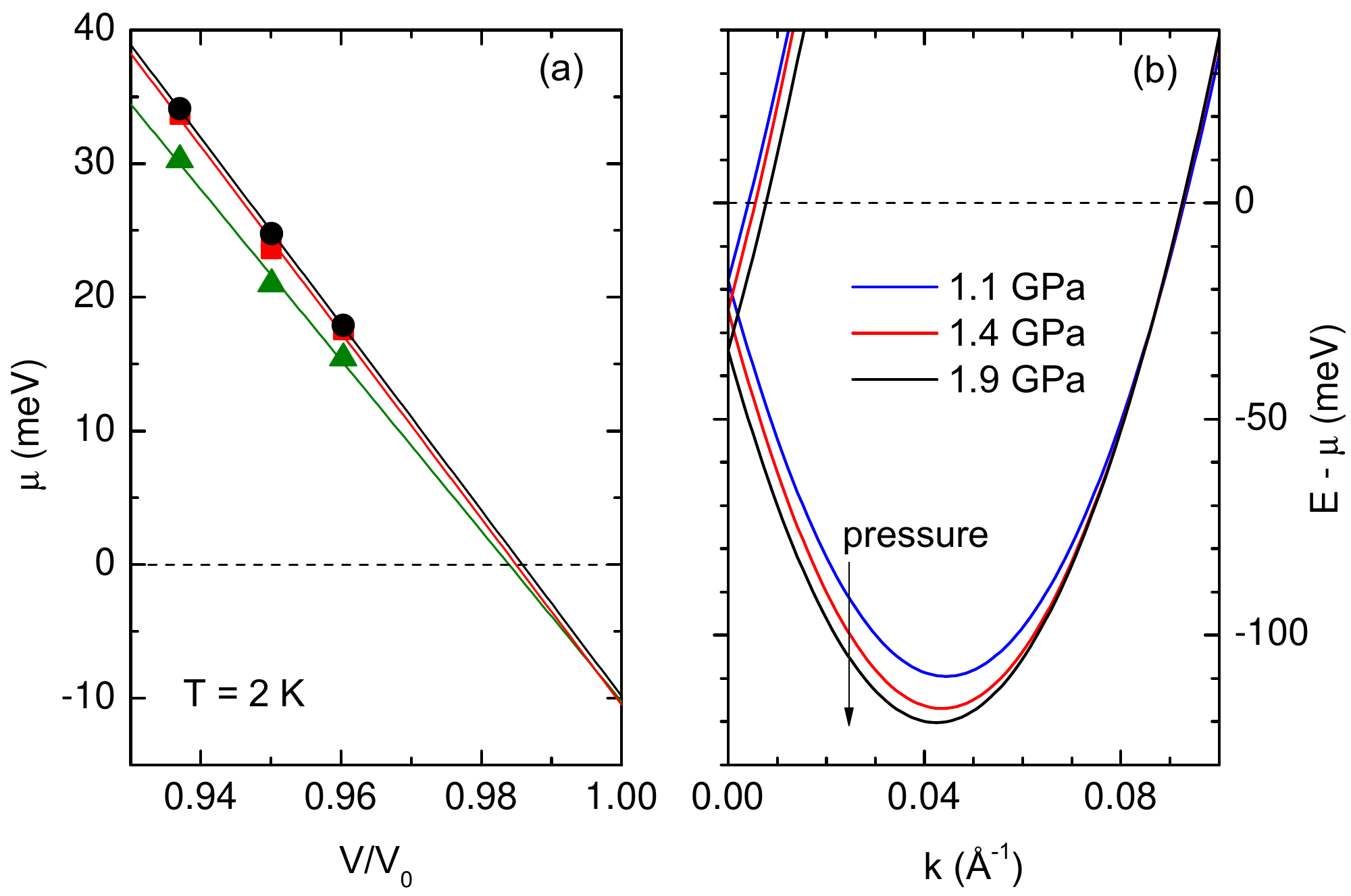}
    \caption{(a) Position of the chemical potential, $\mu$, referenced to the Dirac point.  The black circles indicate values determined by analyzing our data in terms of the Rashba Hamiltonian (Section~\ref{sec:evolution})  The green triangles and red squares indicate values estimated by comparing observed oscillation frequencies to the first principles calculations of Bahramy~\textit{et al.}~\cite{bahramy_2011_1} for BiTeI at ambient pressure and the critical pressure for band inversion, respectively.  Regardless of which estimate is used, extrapolation suggests that the chemical potential lies $\sim 10\,\mathrm{meV}$ below Dirac point at ambient pressure and crosses above it at $V/V_0 = 0.984$, corresponding to a pressure of $0.4\,\mathrm{GPa}$. (b) Band structures at various pressures formed by using the parameters listed in Table~\ref{tab1}.}
 \label{fig5}
\end{figure}

As a consistency check for the method described above, we have also arrived at an estimate of the chemical potential by comparing the observed oscillation frequencies with previous first principles calculations of the band structure.  Bahramy~\textit{et al.}~\cite{bahramy_2011_1} have calculated the band structure for three different relative volumes, corresponding to ambient pressure, and pressures both equal to and above the critical pressure for the band inversion $P_c$.  Our experiments correspond to the pressure range below $P_c$.  We began by digitizing the band structure calculations presented in Fig.~2 of Ref.~\cite{bahramy_2011_1}.  From this data, were were able to determine the specific relationship between the ratio of frequencies of the inner and outer Fermi surface oscillations and the chemical potential (measured from the Dirac point).  The chemical potential versus pressure obtained using this method is presented in Fig.~\ref{fig5}a.  For a given frequency ratio, the estimated $\mu$ is nearly the same regardless of whether the ambient (green triangles) or $P=P_c$ (red squares) calculations are used.  These data also agree quite well with the estimates of the chemical potential arrived at above by analyzing the data in terms of the Rashba Hamiltonian (black circles).

Using $m_1$, $\alpha$ and $\mu$ listed in Table~\ref{tab1} we can arrive at a picture of how the electronic structure (at the bottom of the conduction band) changes with pressure and compare this with the previous predictions.  Fig.~\ref{fig5}b presents the dispersions for the two Rashba sub-bands (all energies are measured from the chemical potential). The various changes under pressure that we had previously discussed are directly evident from Fig.~\ref{fig5}b: the slight decrease in $F^-$, substantial increase in $F^+$, and increase in the chemical potential relative to the Dirac point.  In addition, Fig.~\ref{fig5}b reveals that pressure appears to push the bottom of the conduction band lower towards the valence band.  Although the sign of this trend is consistent with the approach to a pressure induced band inversion, the magnitude of the effect is not in quantitative agreement with predictions.  For the range of relative volumes corresponding to 1.1-1.9 GPa, Ref.~\cite{bahramy_2011_1} (Supplementary material) predicts that the gap should drop by $\sim 70\,\mathrm{meV}$, whereas we find the bottom of the conduction band dropping by only $\sim 10\,\mathrm{meV}$.  However, we note that our model does not include the coupling between the valence and conduction bands, which will become increasingly pronounced as the pressure approaches $P_c$.

\section{Conclusions}
We have carried out a series of measurements of SdH oscillation in BiTeI at pressures up to 1.9 GPa.  This pressure is below the pressure which appears to be required to drive BiTeI through a pressure-induced band inversion~\cite{bahramy_2011_1,xi_2013_1,chen_2013_1}.  An analysis of the results in terms of a simple model Hamiltonian allows an estimate of the pressure dependence of various band structure parameters such as the band mass, $m_1$, the chemical potential $\mu$, and the Rashba parameter $\alpha$.  The estimated value of $\alpha$ is consistent with ARPES measurements~\cite{ishizaka_2011_1}.  Our data suggests a clear increase in the chemical potential from below the `Dirac point' to above it as pressure is increased.  This analysis suggests a lowering of the conduction band minimum with pressure.  The sign of this trend (although not the magnitude) is consistent with the approach to a pressure induced band inversion as predicted in Ref.~\cite{bahramy_2011_1}.  Further studies at higher pressures and in lower carrier density samples will be necessary to explore the possible appearance of novel surface states in the high-pressure non-centrosymmetric topological insulator phase.

\section*{Acknowledgments}
DV and JJH acknowledge the National High Magnetic Field Laboratory's User Collaborative Grants Program (UCGP) for support.  DG and SWT acknowledge support from the Department of Energy (DOE) from grant DOE NNSA DE-NA0001979. SM is a Dirac Post-Doctoral Fellow at the National High Magnetic Field Laboratory, which is supported by the National Science Foundation via Cooperative agreement  No.\ DMR-1157490, the State of Florida, and the U.S. Department of Energy.  DLM  acknowledges support from the National Science Foundation (NSF) from grants NSF DMR-0908029 and NSF DMR-1308972.  We thank Hai-Ping Cheng and David Tanner for informative conversations.

\begin{thebibliography}{10}
\bibitem{rhoxx_note}
The assignment of integer indices to the maxima in $-d^2\rho/dH^2$ (rather than
  the minima) depends on the ratio $\rho_{xx}/\rho_{xy} > 1$. Preliminary Hall
  measurements under pressure on a sample cut from the same parent crystal used
  in this study indicate that this condition is satisfied with
  $\rho_{xx}/\rho_{xy} \geq 1.3$ for fields up to 16 tesla.
\end{thebibliography}
\end{document}